\documentclass{cpbtex}
\usepackage{amssymb,amsmath}
\usepackage{graphicx}
\usepackage{multirow}
\usepackage{hyperref}
\usepackage{appendix}
\usepackage{mathtools}
\usepackage{amsmath}
\usepackage{verbatim}
\usepackage{gensymb}
\usepackage{subfigure}
\usepackage{bm}
\usepackage{threeparttable}
\usepackage{booktabs}
\usepackage{braket}
\usepackage{xcolor}
\usepackage[normalem]{ulem}
\usepackage[mathscr]{euscript}
\usepackage{tabularx}
\usepackage[main=english]{babel}
\addto\captionsenglish{

}

\begin{document}

\title{Inner sublattice competition: a minimal mechanism for reentrant localizations in Aubry-André lattice}

\author{Hangdong Qiu$^{1}$, \ Yunhua Wang$^{1,2,*}$\\
$^1$School of Physical Science and Technology, Lanzhou University, Lanzhou 730000, China\\
$^2$Lanzhou Center of Theoretical Physics,\\
Key Laboratory for Quantum Theory and Applications of the Ministry of Education,\\
Key Laboratory of Theoretical Physics of Gansu Province,\\
Gansu Provincial Research Center for Basic Disciplines of Quantum Physics, \\
Lanzhou 730000, China}

\date{\today}
\maketitle

\begin{abstract}
We construct a double mosaic Aubry-André (DMAA) model with two independent quasiperiodic potentials assigned to corresponding even and odd sublattices of the AA lattice, and show that the competition between them gives rise to multiple localization-delocalization transitions under single-parameter tuning. Such reentrant localizations are verified by comprehensive numerical characterizations including the inverse participation ratio, normalized participation ratio, fractal dimension, and real-space wavefunction distributions. Simulations based on a continuous polariton model yield consistent results, confirming the experimental feasibility of the predicted phenomena. This inner sublattice competition mechanism works in a minimal way for reentrant localizations without additional ingredients or dynamical modifications such as long-range hopping, mixed quasiperiodic and random disorder, or non-Hermitian gain and loss, offering a concise route to engineer reentrant localization based on the canonical AA lattice family.
\end{abstract}

\textbf{Keywords:} %no more than four sets of keywords should be provided
Anderson localization, reentrant localizations, Aubry-André model

\textbf{PACS:} 
72.15.Rn, %Localization effects (Anderson or weak localization)
03.65.Ge %Solutions of wave equations: bound states 

\section{Introduction}
One-dimensional quasiperiodic lattices constitute a fundamental platform for exploring Anderson localization \cite{A1,A2,A3,A4,A5,BEC2008,A6,A7,A8}, many-body localization \cite{Shankarp2013MBL,Michael2015MBL,Xiaopeng2015,Ranjan2015,Lee2017MBL,Dong-LingMBL2017,ALET2018498,
MachineLMBL2018,Shi-XinMBL2018,AbaninREV2019,MBLD2019,Kohlert2019MBL,Liang-Jun2020,LIMBL2021,MBLsharper2024}, and mobility edge physics \cite{DasSarma1988,BiddleME2010,GaneshanME2015,DengRay2019,Hepeng2019,Yucheng2020,AlexME2021,Yunfei2022,QuanLin2022,
XinChi2023,Miguel2023,Longhi2024}, offering highly tunable alternatives to conventional random-disorder systems. The Aubry-André (AA) model provides a canonical example, featuring exact self-duality \cite{Harper1955,Aubry1980,Miguelhd2022}, a solvable single localization transition \cite{Soukoulis1982,Kohmotold1983,Thoulessld1988,Sinha2019ld,Zhihaold2021,g7vd-hgw4}, and intrinsic or extrinsic topological states \cite{Li-Jun2012tp,Yaacovtp2012,KrausYaacovtp2012,Ganeshantp2013,Zhihaotp2013,Huaqingtp2018,Longhitp2019,Jiangtp2019,Zengtp2020}. The mosaic Aubry-André (MAA) model, a sublattice-resolved variant of the AA model, confines the quasiperiodic potential to a single sublattice and produces a drastically different localization landscape: well-defined mobility edges emerge \cite{Yucheng2020}, allowing extended, critical, and localized states to coexist within distinct energy windows. Alternative generalized or modified AA models also allow the mobility edges separating extended and localized states via self-duality \cite{BiddleME2010,GaneshanME2015,DengRay2019,Hepeng2019,Yucheng2020,AlexME2021,Miguel2023,Gopalakrishnan2017,Yanxia2021,Miguelhd2022,
Anomalous2022,Multiple2016,Banerjee2025}, renormalization group method \cite{Renorm2023}, or effective Hamiltonian \cite{effH2025}. Currently, various experimental implementations of different quasiperiodic models, including the AA and MAA models, have been developed to study Anderson localization, such as cold atomic gases \cite{BEC2008,Yunfei2022,H1,H2,H3,H4}, photonic crystals \cite{Yaacovtp2012,Lahini2009,I1,I2,I3,SPME2018,Yi-Jun2025}, acoustic waves \cite{Hefei2008}, etc. In particular, Anderson localization in the AA model is observed directly in cold atomic systems \cite{BEC2008}. The mobility edges are also measured in a MAA lattice implemented via meticulously designed nanophotonic circuits \cite{MAAexp2025}. The stark contrast between these two closely related models highlights that the inner spatial profile of quasiperiodic potentials provides a powerful knob for engineering localization transitions.

Multiple delocalization transitions, namely, reentrant localizations, have recently attracted sustained interest in one-dimensional quasiperiodic systems, as they exhibit novel localization transition behaviors beyond the conventional single transition paradigm and enable multi-step control of localizations or delocalizations \cite{K1,K2,K3,K4,K4.1,K5,K6,K7,K8,K9,K10,K11,K12,K13,K14,K15,K16,K17}. To date, such multiple transitions have been predominantly realized in various one-dimensional quasiperiodic systems including interpolating Aubry-André-Fibonacci (IAAF) model together with its modifications \cite{K1,K2,K3,K4,K4.1}, dimerized lattices with staggered quasiperiodic disorders or on-site potential \cite{K5,K6,K7}, non-Hermitian quasicrystals \cite{K8,K9,K10,K11}, SSH model with disorders \cite{K12,K13,K14}, phase-modulated AA model \cite{K15}, and one-dimensional lattice with periodic and quasiperiodic composite potentials \cite{K16,K17}. Despite these advances, whether such reentrant transitions can emerge solely from amplitude competition between inner sublattice-resolved quasiperiodic potentials remains an open question. Notably, the MAA model naturally possess a sublattice degree of freedom of the AA model. However, the impact of amplitude imbalance between even- and odd-sublattice quasiperiodic potentials on localization transitions, especially across the MAA-to-AA crossover, remains largely unexplored.

\begin{figure}[!htbp]
    \centering
    \includegraphics[width=8 cm]{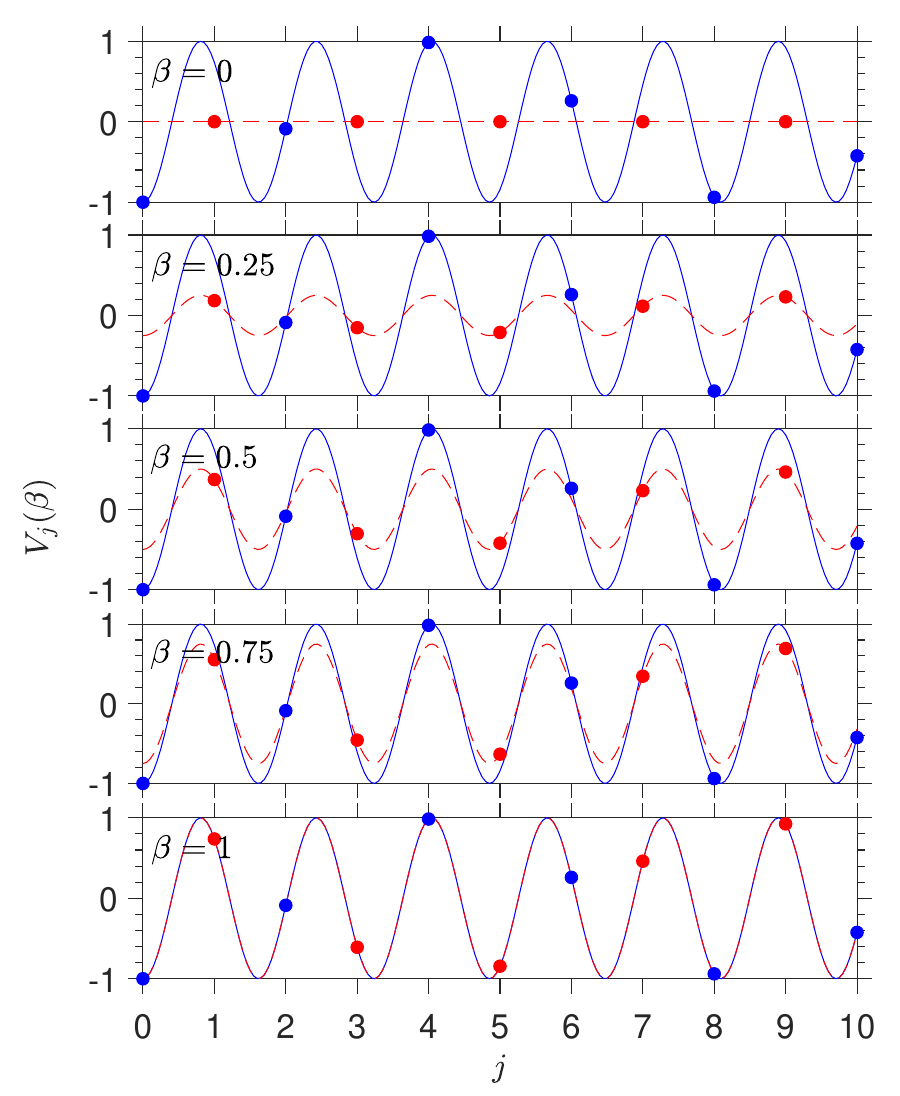}
    \caption{
  Double quasiperiodic potentials $V_j(\beta)$ in Eq.~\eqref{eq:on-site_p} for $\beta = $ 0, 0.25, 0.5, 0.75, and 1, respectively. Blue dots represent the potential on even lattice sites, and red dots represent the potential on odd lattice sites.}
    \label{fig:1}
\end{figure}

Here we demonstrate that such reentrant localization transitions driven by sublattice competition can emerge in a double mosaic Aubry–André (DMAA) lattice. The lattice consists of even and odd sublattices, each hosting an independent quasiperiodic mosaic potential, with a single tunable amplitude parameter $\beta$ controlling their relative strength, as illustrated in Fig.~\ref{fig:1}. Via comprehensive numerical characterization using the fractal dimension Q, inverse participation ratio (IPR), normalized participation ratio (NPR), and real-space wavefunction profiles, we identify three distinct localization regimes as a function of the overall potential strength $\lambda$: (i) Near the critical point of the original AA model ($\lambda =2$), eigenstates in the high-energy window exhibit multiple localization-delocalization transitions as $\beta$ varies, while low-energy states retain AA-like critical behavior; (ii) For relatively weak potentials ($\lambda < 1.6$), low-energy eigenstates remain extended, and high-energy eigenstates support only a single localization transition; and (iii) For relatively strong potentials ($\lambda > 2.4$), most high-energy eigenstates remain localized, while low-energy eigenstates undergo only a single localization transition. Simulations based on a continuous polariton model yield consistent results. This sublattice competition mechanism provides a minimal, previously unidentified route to realize multiple localization transitions based on the AA model. Our work establishes sublattice-resolved potential design as a concise strategy for engineering localization behavior in the canonical AA lattice family.

\section{The DMAA model and methods}

We consider a one-dimensional chain with $2N$ sites in the presence of double quasiperiodic on-site potentials described by the following Hamiltonian 
\begin{equation}
H = t \sum_j \left( \hat{c}_j^\dagger \hat{c}_{j+1} + h.c. \right) + \lambda \sum_j V_j(\beta) \hat{c}_j^\dagger \hat{c}_j,
\label{eq:h}
\end{equation}
where $\hat{c}_j^\dagger$ and $\hat{c}_j$  are the creation and annihilation operators at site $j$, respectively, $t$ is the nearest-neighbor hopping and assumed to be $1$ as the energy unit, and $\lambda$ is the strength of the quasiperiodic on-site potential $V_j(\beta)$ defined as
\begin{equation}
V_j(\beta) = 
  -\cos(2\pi\alpha j) \cdot \delta_{j,2m} - \beta \cos(2\pi\alpha j) \cdot \delta_{j,2m+1}.
\label{eq:on-site_p}
\end{equation}
Here, $m$ is an integer from 0 to $N$, $\alpha = \frac{\sqrt{5} - 1}{2}$ denotes the spatial modulation frequency, and $\beta$ is the tuning parameter ranging from 0 to 1. Eq.~\eqref{eq:on-site_p} indicates that the even sites have the potential $\cos(2\pi\alpha j)\delta_{j,2m}$ with $j=2m$, and the odd sites have the potential $\beta  \cos(2\pi\alpha j)\delta_{j,2m+1}$ with the tuning parameter $\beta$ and $j=2m+1$. This means that the one-dimensional chain has double quasiperiodic modulated potentials. In the case of $\beta = 0$, Eq.~\eqref{eq:on-site_p} gives the MAA model, and $\beta = 1$ corresponds to the standard AA model. For the other cases with $\beta$ varying from 0 to 1, Eq.~\eqref{eq:on-site_p} shows a continuous evolution from MAA to AA models, as illustrated in Fig.~\ref{fig:1}.

\begin{figure*}[!htbp]
    \centering
    \includegraphics[width=16 cm]{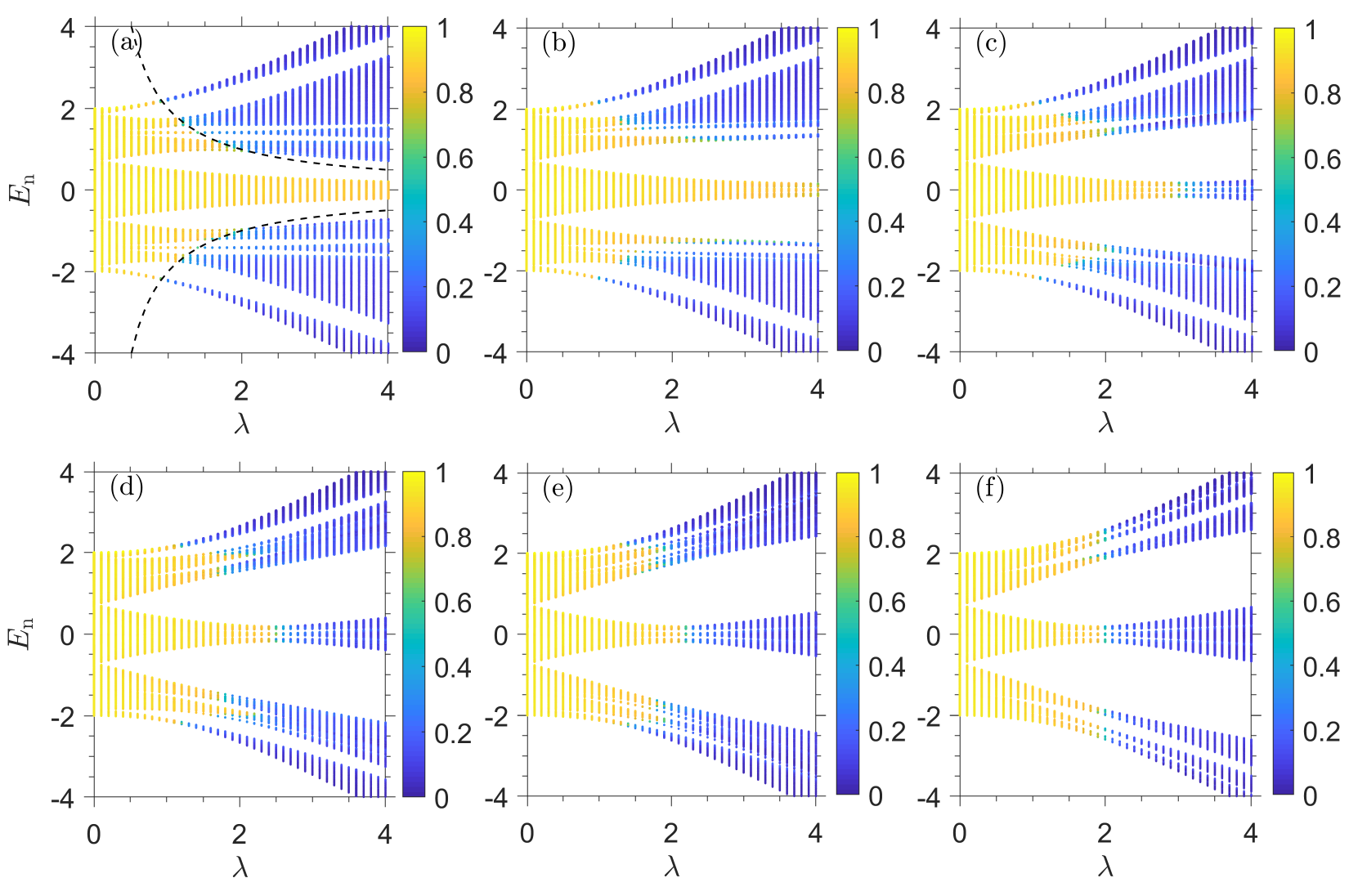}
    \caption{
  (a)-(f) Fractal dimension $Q$ of eigenstates as a function of eigenenergy $E_n$ and quasiperiodic potential strength $\lambda$ for $\beta=$ 0, 0.2, 0.4, 0.6, 0.8, and 1, respectively, with system size $L = 1220$. The two black dashed lines in panel (a) represent the mobility edges of MAA model.}
    \label{fig:2}
\end{figure*}

The IPR and NPR are two significant diagnostic quantities characterizing the localization transition \cite{K5,S1,S2}. The IPR for the $n$th eigenstate $\psi^{(n)}$ of the system is defined as
 \begin{equation}
  \text{IPR}^{(n)} = \sum_{j=1}^L \left| \psi_j^{(n)} \right|^4, 
  \label{eq:IPR}
\end{equation} 
where $L=2N$ as the system size, $j$ denotes the $j$th lattice site, and the wavefunction $\psi^n$ satisfies the normalization condition. For an extended state, the IPR satisfies $\text{IPR} \to 1/L$, and drops to 0 in the thermodynamic limit. For a localized state the IPR takes a finite value that is independent of $L$. The NPR for the $n$th eigenstate of the system is defined as  
\begin{equation}
   \text{NPR}^{(n)} = \left( L \sum_{j=1}^L \left| \psi_j^{n} \right|^4 \right)^{-1}.
   \label{eq:NPR}
\end{equation}
For an extended state, the NPR takes a finite value, and for a localized state the NPR tends to zero. The fractal dimension Q \cite{A5,Yucheng2020} is defined as
 \begin{equation}
  Q = -\lim_{L \to \infty} \frac{\ln \text{IPR}}{\ln L}. 
  \label{eq:frac}
 \end{equation}
Eq.~\eqref{eq:frac} indicates that $\text{Q} \to 0$ represents a localized state, and $\text{Q} \to 1$ corresponds to an extended state.

\section{Results and discussion}
Figure \ref{fig:2} maps the full evolution of eigenstate localization across the energy spectrum as the DMAA model continuously interpolates from the pure MAA limit ($\beta = 0$) to the pure AA limit ($\beta = 1$). At $\beta = 0$ in Fig.~\ref{fig:2}(a), the system reproduces the known MAA behavior with clearly identifiable mobility edges $E_c=\pm 2/\lambda$ separating extended, critical, and localized states \cite{Yucheng2020}. As $\beta$ increases, the mobility edges evolve continuously but complexly, until all mobility edges eventually vanish upon approaching the AA limit \cite{Aubry1980,T1}, as shown in Fig.~\ref{fig:2}(f). Owing to the interplay of two distinct quasiperiodic potentials residing on separate sublattices, an exact analytical description of these mobility edge trajectories remains nontrivial and is not pursued here. Instead, we draw attention to the prominent localization transition features of all eigenstates with $\beta$ for different $\lambda$, especially for these eigenstates adjacent to the original MAA mobility edge near the transition point $\lambda = 2$ of the AA model. In the following sections, we characterize localization transitions in quantitative detail using the IPR, NPR, $Q$ and real-space wavefunction distributions.

\begin{figure}[!htbp]
    \centering
\includegraphics[width=16 cm]{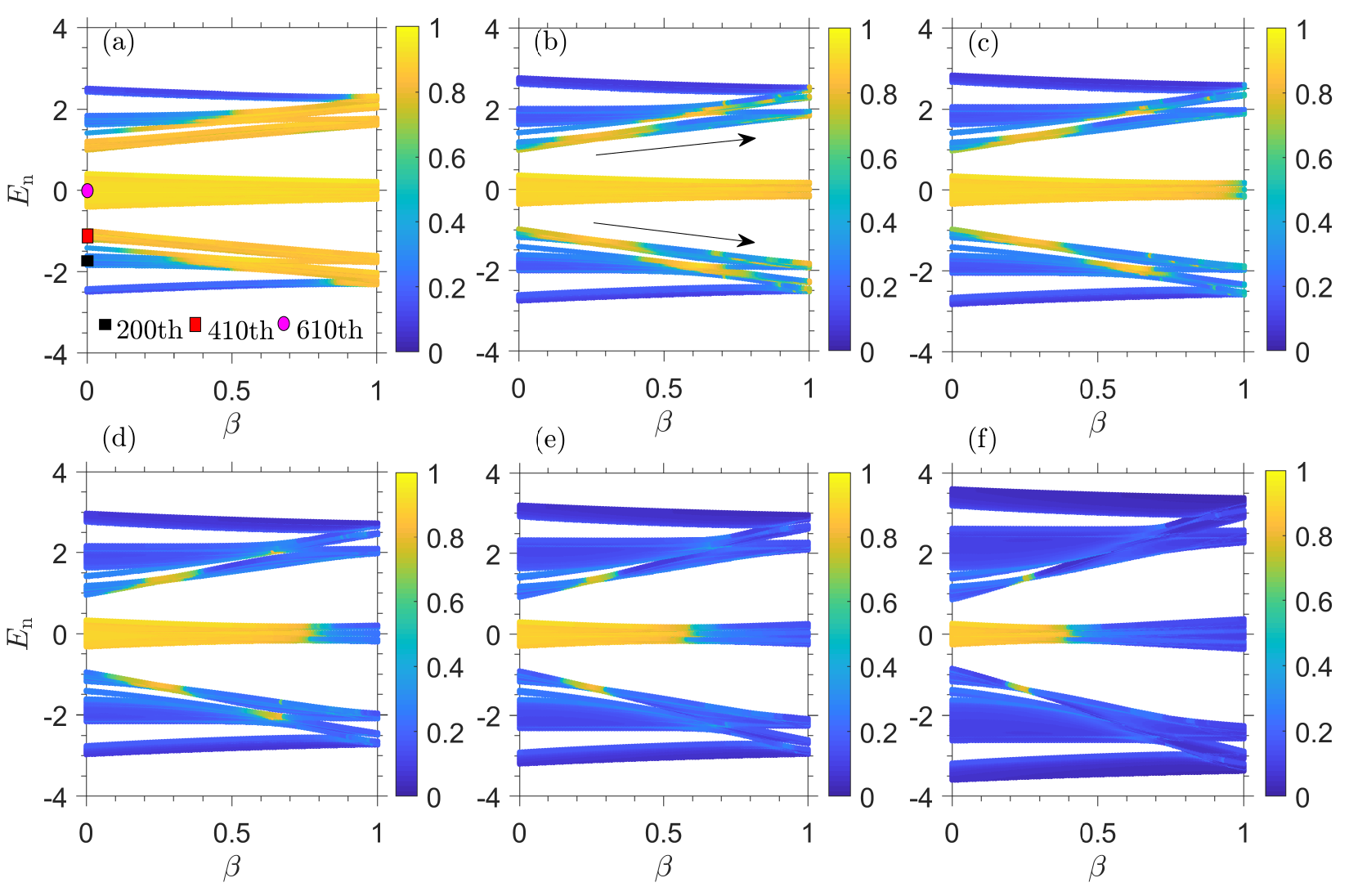}
    \caption{
  (a)-(f) Fractal dimension $Q$ of all eigenstates as a function of $E_n$ and $\beta$ for $\lambda=$ 1.5, 1.9, 2, 2.2, 2.5, and 3, respectively, with system size $L = 1220$. Along the arrows in (b), these eigenstates undergo multiple localization transitions with increasing $\beta$.}
    \label{fig:3}
\end{figure}

We specify different $\lambda$ to explore how the localization transition evolves with $\beta$ ranging from 0 to 1 in the DMAA structure, namely the crossover from the MAA to the AA model. Here we refer the relatively high energy to the energy with about $|E_n| > 0.5$, since the negative and positive signs represent different types of quasi-particles. We divide $\lambda$ into three cases, (i) smaller $\lambda$, namely, with about $\lambda < 1.6$, (ii) moderate $\lambda$, with its value near the transition point $\lambda = 2$, and (iii) larger $\lambda$, namely, with about $\lambda > 2.4$. We now discuss the localization transitions for the three cases. In Figs.~\ref{fig:3}(a) - \ref{fig:3}(f), the fractal dimension $Q$ of all eigenstates is plotted as a function of $E_n$ and $\beta$ for $\lambda=$ 1.5, 1.9, 2, 2.2, 2.5, 3, respectively. Firstly, for smaller $\lambda$ with $\lambda = 1.5$ as an example in Fig.~\ref{fig:3}(a), these eigenstates near zero energy with about $|E_n| < 1.33$ keep extended, and the other eigenstates with higher energy undergo a localization transition \textit{once}. Secondly, near the transition point $\lambda = 2$ of the AA model with $\lambda = 1.9$, 2, and 2.2 in Figs.~\ref{fig:3}(b) - \ref{fig:3}(d), with increasing $\beta$, besides single localization transition, some eigenstates with high energy obviously undergo \textit{multiple} localization transitions, as denoted by the arrows. These eigenstates near zero energy can be extended for $\lambda < 2$ in Fig.~\ref{fig:3}(b), critical for $\lambda = 2$ in Fig.~\ref{fig:3}(c), or localized for $\lambda > 2$ in Fig.~\ref{fig:3}(d), respectively. Thirdly, for larger $\lambda$ with $\lambda = 2.5$ and $3$ in Fig.~\ref{fig:3}(e) and Fig.~\ref{fig:3}(f), respectively, these eigenstates near zero energy only undergo single localization transition, and a few eigenstates with higher energy experience transitions from localized states to extended states and finally back to localized states, for $\beta$ inside about $(0.1, 0.4)$.

\begin{figure*}[!htbp]
    \centering   
\includegraphics[width=16 cm]{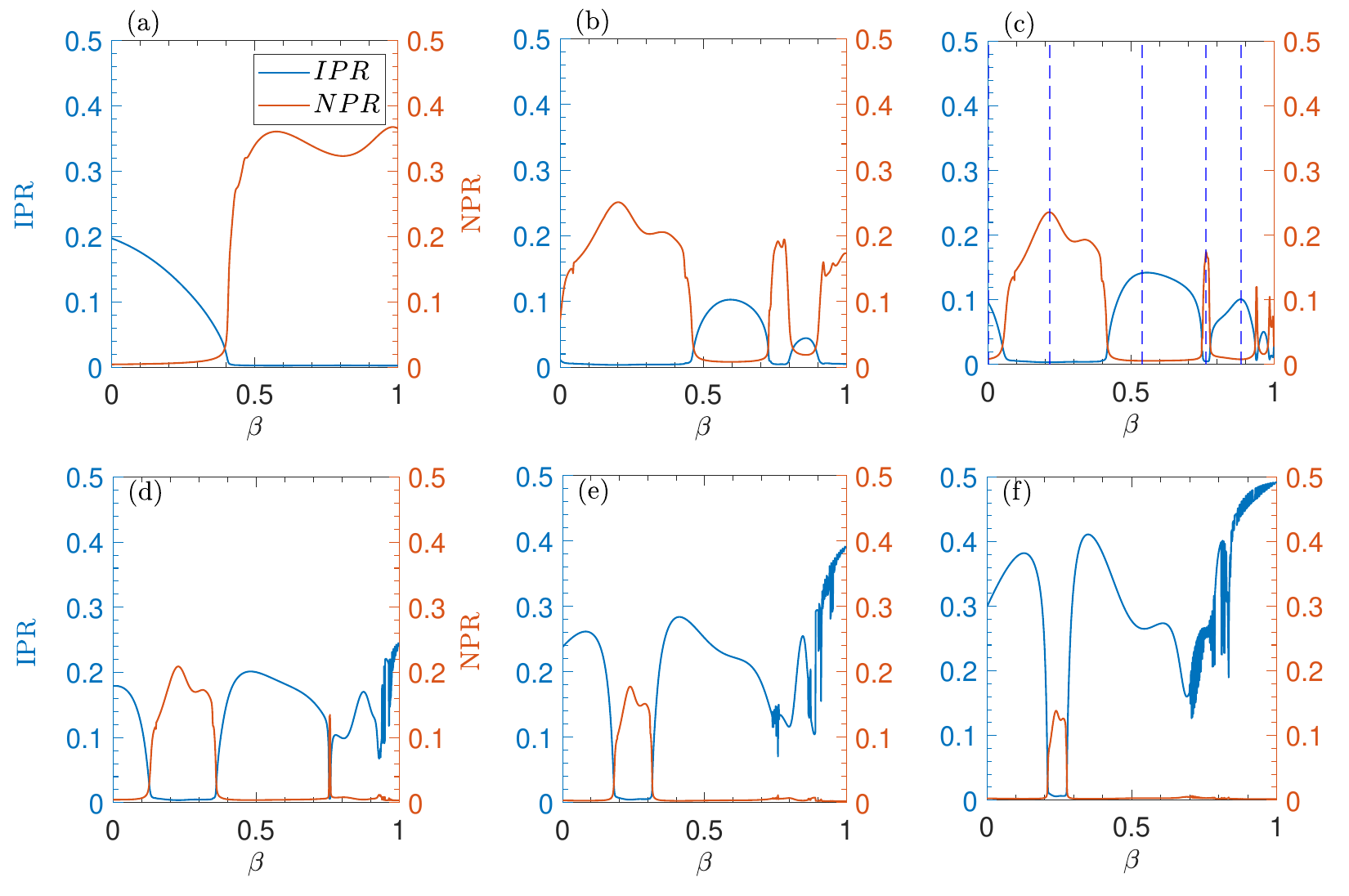}
    \caption{
    The IPR and NPR as functions of $\beta$ in (a) for the $200$th eigenstate in Fig.~\ref{fig:3}(a) with $\lambda=$ 1.5, and in (b)-(f) for the $410$th eigenstate in Figs.~\ref{fig:3}(b)-~\ref{fig:3}(f) with $\lambda=$ 1.9, 2, 2.2, 2.5, and 3, respectively. These vertical dashed lines in (c) specify $\beta =$ 0, 0.215, 0.538, 0.761, and 0.884, where the wavefunctions of the  $410$th eigenstate are correspondingly plotted in Fig.~\ref{fig:5}.}
    \label{fig:4}
\end{figure*}

We select some specified eigenstates for the above three cases of the localization transitions and illustrate how the IPR and NPR for high-energy eigenstates vary with $\beta$, so as to more clearly explore these localization transition features from MAA to AA models in the DMAA model. For the first case with smaller $\lambda= 1.5$, both IPR and NPR of the eigenstate with higher energy such as the $200$th one here in Fig.~\ref{fig:4}(a) exhibit an obvious abrupt change with increasing $\beta$, manifesting one localization transition. For the second case with $\lambda$ near $2$, the IPR and NPR of some high-energy eigenstates, e.g., the $410$th state, here in Figs.~\ref{fig:4}(b) - \ref{fig:4}(d) show multiple abrupt changes from zero to a finite value or vice versa. This indicates that such high-energy eigenstates undergo multiple localization transitions rather than single transition from MAA to AA models. For the third case with larger $\lambda$ in Figs.~\ref{fig:4}(e) and~\ref{fig:4}(f), we can see that the IPR of some high-energy eigenstates, e.g., the $410$th state, displays a remarkably abrupt change from a finite value to zero and then to a finite value, and vice versa for NPR. This means such eigenstates evolve from localized to extended and back to localized states.

\begin{figure*}[!htbp]
\centering
\includegraphics[width=16 cm]{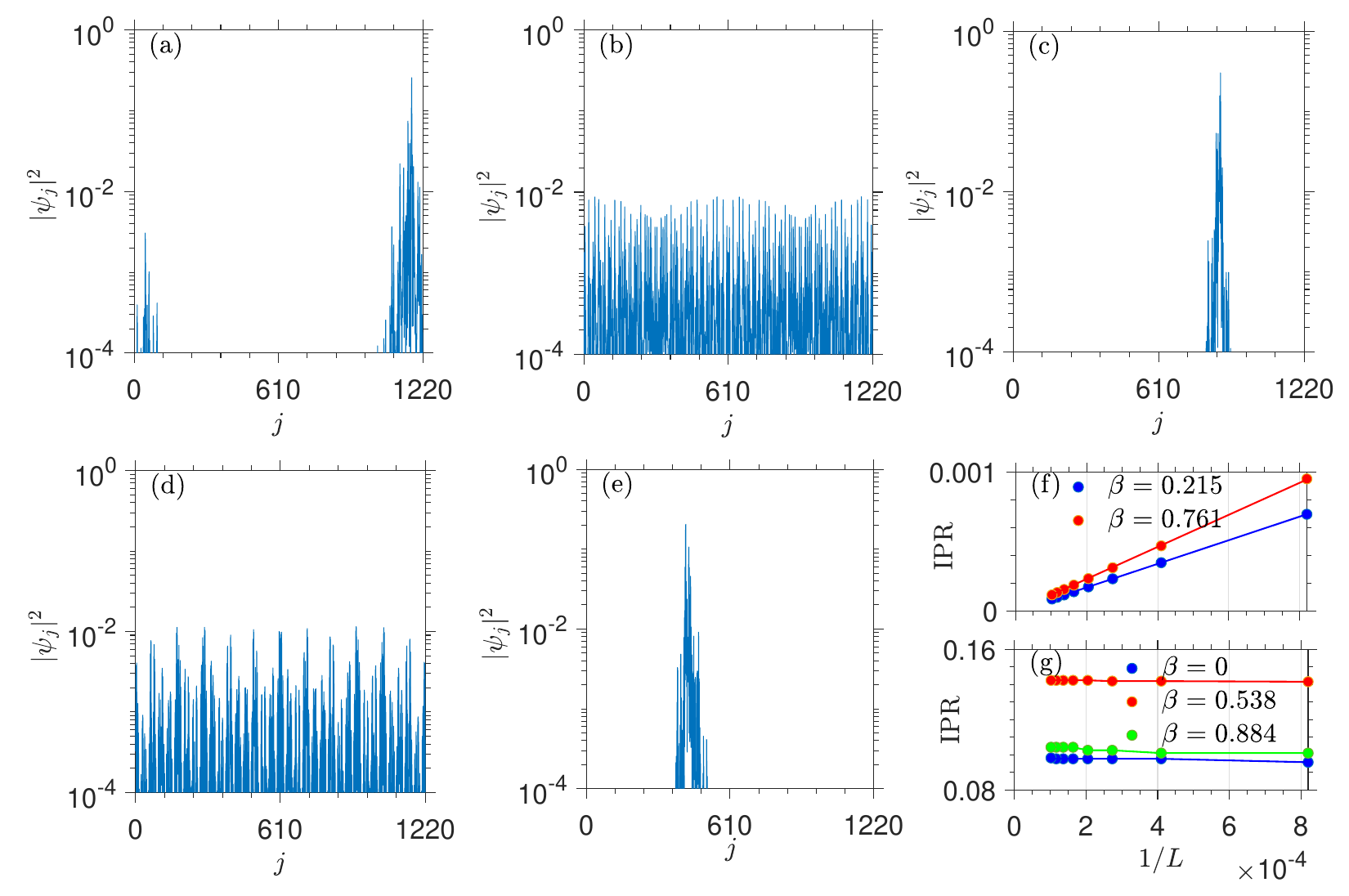}
\caption{(a)-(e) Spatial distribution of the wave function for the 410th eigenstate of the DMAA model with $\lambda= 2$ for $\beta=0,0.215,0.538,0.761$, and $0.884$, respectively, where the vertical axis is plotted on a logarithmic scale.
(f) and (g) Dependence of the IPR on system length for corresponding $\beta$.}
\label{fig:5}
\end{figure*}

To confirm the above extended and localized states and their localization transitions characterized by the IPR and NPR, we further calculate their real-space wavefunctions. Here, the localization transition for $\lambda = 2$ in Fig.~\ref{fig:4}(c) is chosen as an example, and the other case is similar. We compute and plot the wavefunctions for five different $\beta$ in Fig.~\ref{fig:4}(c). For zero NPR (a nonzero finite IPR), e.g., for $\beta = 0$ in Fig.~\ref{fig:5}(a), $\beta = 0.538$ in Fig.~\ref{fig:5}(c), and $\beta = 0.884$ in Fig.~\ref{fig:5}(e), their wavefunctions are highly localized at few positions and hence manifest their localized state features. For zero IPR, e.g., for $\beta = 0.215$ in Fig.~\ref{fig:5}(b), and $\beta = 0.761$ in Fig.~\ref{fig:5}(d), their wavefunctions are highly extended and hence exhibit their extended state features. We also discuss the size dependence behaviors of these extended and localized states. As shown in Fig.~\ref{fig:5}(f), the IPR of extended states for $\beta = 0.215$ and $0.761$ is inversely proportional to $L$, and with $L \to \infty$ the IPR is close to zero in the thermodynamic limit. The IPR of localized states for $\beta = 0$, 0.538 and 0.884 is independent of $L$, and with increasing $L$ the IPR is almost constant with a nonzero finite value in Fig.~\ref{fig:5}(g).

\begin{figure*}[!htbp]
    \centering
    \includegraphics[width=16 cm]{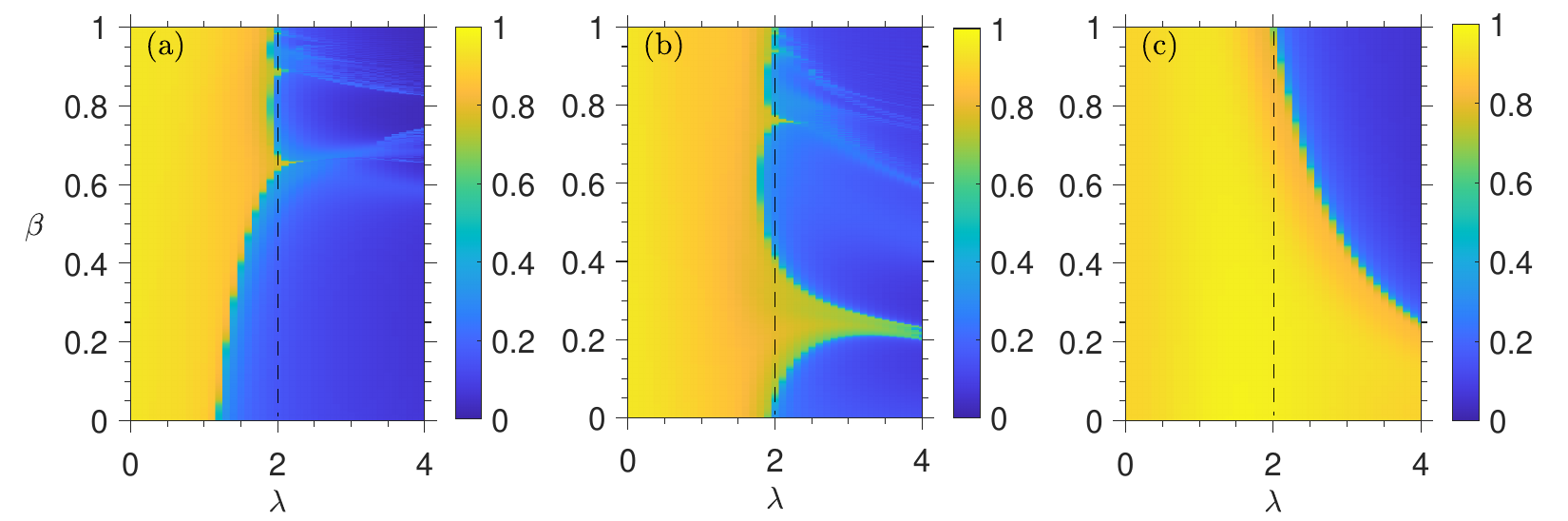}
   \caption{
(a)-(c) The phase diagram characterized by fractal dimension $Q$ in the ($\lambda$, $\beta$) plane for the $200$th, $410$th, and $610$th eigenstates, respectively, in DMAA model. 
}
    \label{fig:6}
\end{figure*}

Next, we summarize our main findings on the localization transition features from the MAA to AA models via the phase diagrams characterized by the fractal dimension Q with respect to $\beta$ and $\lambda$, as shown in Figs.~\ref{fig:6}(a) - \ref{fig:6}(c). We select the 200th and 410th eigenstates to discuss localization transitions of high-energy eigenstates, and take the 610th eigenstate representing the low-energy states near zero energy. The fractal dimension Q of the 200th, 410th and 610th eigenstates as a function of $\beta$ and $\lambda$ are shown in Figs.~\ref{fig:6}(a) - \ref{fig:6}(c), respectively.  For high-energy eigenstates, a single localization transition can emerge but only at $\lambda<2$ as shown in Fig.~\ref{fig:6}(a), multiple localization transitions occur around $\lambda=2$ in Figs.~\ref{fig:6}(a) and \ref{fig:6}(b), and twice localization transitions occur at about $\lambda>2.5$ in Figs.~\ref{fig:6}(b). For low-energy eigenstates around zero energy in Fig.~\ref{fig:6}(c), they almost stay extended for $\lambda<2$, and change from extended to critical states at $\lambda=2$, and undergo only a single localization transition for $\lambda>2$. Moreover, the transition takes place at smaller $\beta$ as $\lambda$ increases.

We finally discuss the physical origin of these peculiar multiple localization-transition behaviors in the DMAA model. It is well known that distinct disorder configurations at fixed hopping amplitudes can yield different distributions of extended, critical, and localized eigenstates as functions of energy and disorder strength. Previous studies have confirmed that competitions between dimerization and quasiperiodic disorder \cite{K5,K7}, or between periodic and quasiperiodic potentials \cite{K16,K17}, give rise to reentrant localization. Our DMAA model in Eq.~\eqref{eq:on-site_p} can be viewed as a superposition of two modified MAA models: one with on-site potentials acting exclusively on even lattice sites, and the other with on-site potentials restricted solely to odd lattice sites. The superposition of these two quasiperiodic potentials induces interplays and competitions among extended, localized, and critical states, which underpin the multiple localization–delocalization transition features of the DMAA model. We therefore reveal that competition between the two sublattice-confined quasiperiodic potentials provides a minimal mechanism for generating multiple reentrant localization–delocalization transitions in one-dimensional quasiperiodic systems.

\begin{figure*}[!htbp]
    \centering
    \includegraphics[width=16 cm]{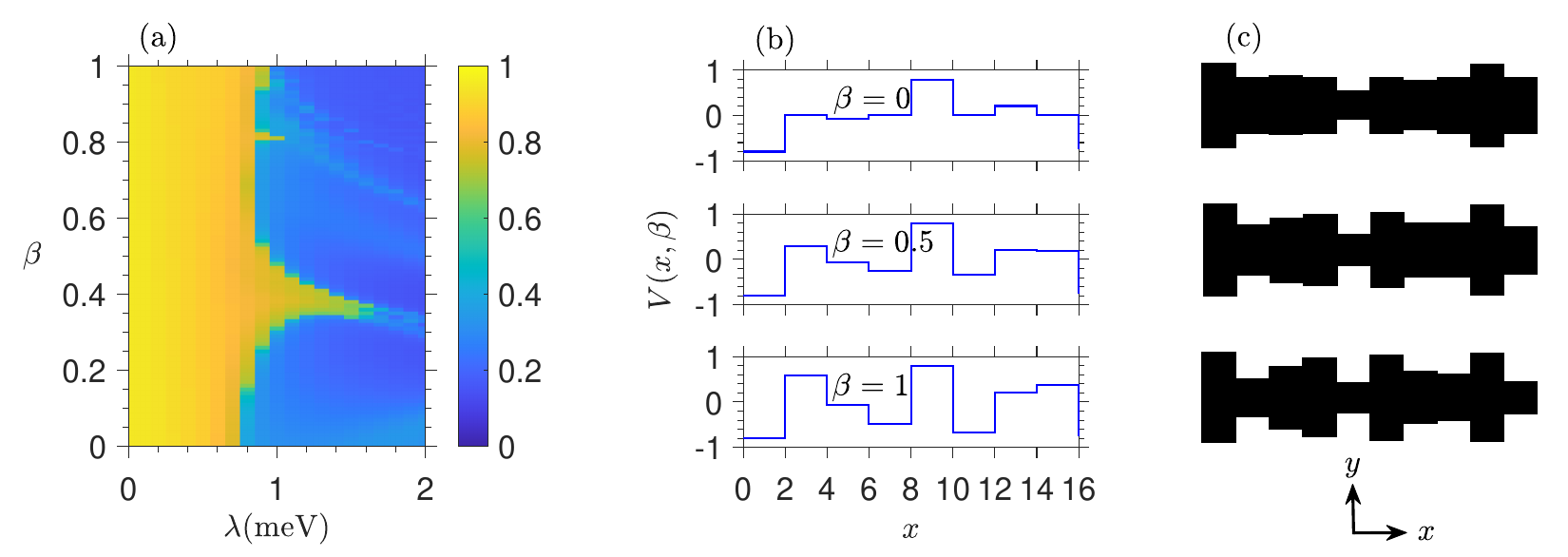}
    \caption{
  (a) Fractal dimension $Q$ of the 410th eigenmode of the continuum DMAA model as a function of $\lambda$ and $\beta$, where
  the system size is $L = 2440$ $\mu$m with a pitch of $a = 2$ $\mu$m.
  (b) Potential functions corresponding to different values of $\beta$ in the continuum DMAA model.
  (c) Top view of polariton structures for different $\beta$ values.
  In (b) and (c), $\beta$ is set to 0, 0.5 and 1, respectively.
}
\label{fig:7}
\end{figure*}

\section{Continuum model simulations on cavity polaritons}

The state-of-the-art electron-beam lithography combined with dry etching is applied to prepare laterally modulated quasi-1D photonic wires and observe the transport, localization and phase transition of cavity polariton modes, including the characteristic features  of fractal energy spectrum in Fibonacci model \cite{U1} and the band-selective multiple
localization-delocalization transitions in the IAAF model \cite{K1}. The experimental polariton modes modulated by the quasiperiodic IAAF potential can also be simulated by continuum Hamiltonian model \cite{K1}. Following the simulation scheme, we construct a continuum version of the DMAA model, 
\begin{equation}
  H\psi(x) = \left[ -\frac{\hbar^2}{2m}\nabla^2 + \lambda V(x,\beta) \right]\psi(x)  
  \label{eq:q6}
\end{equation}
where $m$ is the mass of polariton, and $V(x,\beta)$ is a step function derived from the modulation potential $V_j(\beta)$ in Eq.~(\ref{eq:on-site_p}) with steps of length $a$ (here, $a = 2$ $\mu$m) and the effective strength $\lambda$ of quasiperiodic potential. For comparisons on that of tight-binding model in Fig.~\ref{fig:6}(b), we use Eq.~(\ref{eq:q6}) calculate the fractal dimension $Q$ of the 410th eigenmode as a function of $\beta$ and $\lambda$ and plot the results in Fig.~\ref{fig:7}(a). As we can see, the continuum DMAA model exhibits similar multiple localization–delocalization transition behaviors as that of tight-binding DMAA model. This means that our theoretically predicted localization transition features of the DMAA model can be observed in the  cavity polariton experiments.

\section{Conclusion}

In summary, we have constructed a double mosaic Aubry–André (DMAA) model that enables a continuous crossover from the MAA limit to the AA limit via a single tunable amplitude parameter $\beta$. Through comprehensive numerical characterizations including the inverse participation ratio, normalized participation ratio, fractal dimension, and real-space wavefunction profiles, we demonstrate that eigenstates in a characteristic energy window exhibit multiple reentrant localization-delocalization transitions as $\beta$ is varied, in contrast to the single full-spectrum transition in the conventional AA model.

The physical origin of this behavior lies in the competition between two independent quasiperiodic potentials residing on the even and odd sublattices. This sublattice competition mechanism provides a minimal and previously unreported route to realize multiple localization transitions within the canonical nearest-neighbor tight-binding framework, without requiring multiband structures, long-range hopping, random disorder, or non-Hermitian gain and loss.

We further confirm that the same reentrant localization behavior can be reproduced in a continuous polariton platform, indicating that the observed phenomena are not artifacts of the discrete lattice model but generic physical effects accessible in realistic experimental systems. Our work establishes sublattice-resolved potential engineering as a straightforward strategy for tailoring localization properties in quasiperiodic systems. Looking forward, this sublattice competition framework can be readily extended to non-Hermitian quasiperiodic systems and interacting many-body localization phases.

\section*{Data availability statement}
All data needed to evaluate the conclusions in the paper are present in the paper. Additional data related to this paper may be requested from the corresponding author.

\addcontentsline{toc}{chapter}{Acknowledgment}
\section*{Acknowledgment}
This work was supported by the National Natural Science Foundation of China (Grant No. 12247101), the Fundamental Research Funds for the Central Universities (Grant No. lzujbky-2025-jdzx07), the Natural Science Foundation of Gansu Province (No. 22JR5RA389, No. 25JRRA799), and the 111 Project (Grant No. B20063).

\bibliographystyle{iopart-num} 
\bibliography{apssamp}

\end{document}